\documentclass[doublecol]{epl2}

\usepackage{amssymb}
\usepackage{amsfonts}
\newcommand{\dd}{\mathrm{d}}

\title{Fluctuations of the total entropy production in stochastic systems}
\author{S. Joubaud, N. B. Garnier, S. Ciliberto}
\institute{
	Universit\'e de Lyon\\
	Laboratoire de Physique de l'ENS Lyon, CNRS UMR 5672,
        46, All\'ee d'Italie, 69364 Lyon CEDEX 07, France}
\date{\today}

\abstract{
Fluctuations of the total entropy are experimentally investigated in two  stochastic systems in a non-equilibrium steady state  : an electric circuit with an imposed mean current and a harmonic oscillator driven out of equilibrium by a periodic torque. In these two linear systems, we study the total entropy production which is the entropy created to maintain the system in the non-equilibrium steady state. Fluctuation theorem holds for the total entropy production in the two experimental systems, both for all observation times and for all fluctuation magnitudes.}

\begin{document}

\maketitle
Thermodynamics of systems at equilibrium has been developed by defining the internal energy, the injected work, and the dissipated heat.
The first law of Thermodynamics describes energy conservation and gives a relation between those three energies. The second law of Thermodynamics
imposes that the entropy variation is positive for a closed system. Statistical Physics further gives a microscopic definition of entropy,
which allows analytical results on entropy production. The extension of Thermodynamics to systems in non-equilibrium steady states
is an active field of research. Within this context, the first law has been extended for stochastic systems described by a Langevin 
equation~\cite{Sekimoto1998,Blickle2006,Joubaud2007}. It has been noted that the second law is not verified at all times 
but only in average, {\em i.e.} over macroscopic times : entropy production can have instantaneously negative values. The probabilities of getting 
positive and negative entropy production are quantitatively related in non-equilibrium systems by the Fluctuation Theorem (FT). This theorem has been 
first demonstrated in deterministic systems~\cite{Evans1993,GallavottiCohen95,EvansSearles} and secondly extended to stochastic dynamics~\cite{Kurchan1998, Lebowitz1999, Harris2006, Chetrite2007}. 
FTs for work and heat fluctuations have been theoretically and experimentally studied in Brownian systems described by the Langevin 
equation~\cite{Farago2002, VanZon-Cohen, VanZon-Cohenbis, Wangetal, Garnier2005, Imparato2007, Douarche2006, Joubaud2007, Blickle2006, Taniguchi}. 
For these systems in non-equilibrium steady state, Fluctuation Theorems hold only in the limit of infinite time:
\begin{equation}
\Phi(a) \equiv \ln\left( \frac{p(X_\tau = +a)}{p(X_\tau=-a)}\right) \; \rightarrow \; \frac{a}{k_BT}\; \rm{for} \;\tau \rightarrow \infty
\label{FT:def} 
\end{equation}
where $k_B$ is the Boltzmann constant, $T$ the temperature of the heat bath and $p(X_\tau)$ is the probability density function (PDF) of $X_\tau$. $\Phi$ is called symmetry function and $X_\tau$ stands for either injected work or dissipated heat, averaged over a time lag $\tau$. For the injected work, eq.~\ref{FT:def} is valid for all fluctuation magnitudes $a$ ; for the dissipated heat, eq.~\ref{FT:def} is satisfied for values lower than the mean value~\cite{VanZon-Cohenbis}.

We are interested here in the total entropy production in a non-equilibrium steady state (NESS), introduced in ref.~\cite{Seifert2005, Speck2005, Seifert2007} and directly related to previous work on housekeeping heat~\cite{Oono1998, Hatano2001}. Entropy production has been studied both theoretically and experimentally in several systems~\cite{Trepagnier2004, Andrieux2007, Cleuren2006, Tietz2006, Imparato2005, Gomez2006, Speck2007}. 
In~\cite{Seifert2005}, Seifert and Speck have shown that 
the total entropy production for a stochastic system described by a first order Langevin 
equation in a NESS satisfies a Detailed Fluctuation Theorem (DFT) :
\begin{equation}
\Phi(a) = \ln\left( \frac{p(X_\tau = +a)}{p(X_\tau=-a)}\right) = \frac{a}{k_BT}\quad \forall \tau  \quad \forall a
\label{eq:FT_total_entropy}
\end{equation}
The relation (\ref{eq:FT_total_entropy}) is valid for all integration time $\tau$ and all fluctuation magnitudes of the total entropy production. This relation is closely 
related to the Jarzynski and Crooks non-equilibrium work relations~\cite{Jarzynski, Crooks} 
which can be exploited to measure equilibrium free energy in 
experiments~\cite{Douarche05, Collin, Ritort, Hummer, Park} and it has been extended to Markov processes~\cite{Puglisi2006}.

In this Letter, we measure the total entropy production in two out-of-equilibrium systems 
and show that this quantity verifies eq.~(\ref{eq:FT_total_entropy}). In the first part 
of the letter, we recall the general definition of dissipated heat, 
"trajectory-dependent" entropy and total entropy production. In the second part, we detail 
our two experimental systems. The first one is an electric circuit maintained 
in a NESS by an injected mean current. The second system is a 
torsion pendulum driven in a NESS by forcing 
it with a periodic torque.

The heat dissipated by the system is the heat given to the thermostat during a time $\tau$ ;
we note it $Q_\tau$. It is related to 
the work $W_{\tau}$, given to the system during the time $\tau$, and 
to the variation of internal energy $\Delta_\tau U$ during this period,
thanks to the first law of Thermodynamics: 
\begin{equation}
Q_\tau = W_{\tau} - \Delta_{\tau} U \,.
\label{eq:first_law}
\end{equation}
Expressions of $W_{\tau}$ and $\Delta_{\tau} U$ for our two 
experimental setups are given below. 
Following notations of ref~\cite{Seifert2005}, we define the entropy variation in the system during a time $\tau$ as :
\begin{equation}
\Delta s_{\rm{m},\tau} = \frac{1}{T}Q_\tau
\label{eq:def_medium_entropy}
\end{equation}
where $T$ is the temperature of the heat bath. For thermostated systems, entropy change in medium behaves like the dissipated heat. The non-equilibrium Gibbs entropy is :
\begin{equation}
S(t) = -k_B \int \dd \vec{x} p(\vec{x}(t),t,\lambda_t) \ln p(\vec{x}(t),t,\lambda_t) = \langle s(t) \rangle
\end{equation}
where $\lambda_t$ denotes the set of control parameters at time $t$ 
and $p(\vec{x}(t),t,\lambda_t)$ is the probability density function to find the particle at 
a position $\vec{x}(t)$ at time $t$, for the state corresponding to $\lambda_t$. 
This expression allows the definition of a "trajectory-dependent" entropy :
\begin{equation}
s(t) \equiv -k_B \ln p(\vec{x}(t),t,\lambda_t)
\label{eq:def_Gibbs_entropy}
\end{equation}
The variation $\Delta s_{\rm{tot},\tau}$ of the total entropy $s_{\rm{tot}}$ during a time $\tau$ is the sum of the entropy change in the system during $\tau$ and the variation of the "trajectory-dependent" entropy in a time $\tau$, $\Delta s_\tau\equiv s(t+\tau) - s(t)$ :
\begin{equation}
\Delta s_{\rm{tot},\tau} \equiv s_{\rm{tot}}(t+\tau) - s_{\rm{tot}}(t) = \Delta s_{\rm{m},\tau} + \Delta s_\tau
\label{eq:def_total_entropy}
\end{equation}
In this letter, we study fluctuations of $\Delta s_{\rm{tot},\tau}$ computed using (\ref{eq:def_medium_entropy}) and (\ref{eq:def_Gibbs_entropy}). We will show that $\Delta s_{\rm{tot},\tau}$ satisfies a DFT (eq.~(\ref{eq:FT_total_entropy})). In ref.~\cite{Puglisi2006}, the relevance of boundary terms like $\Delta s_\tau$ has already been pointed out.

Our first experimental system is an electric circuit composed of a resistance $R = 9.22$~M$\Omega$ in parallel with a capacitance $C = 278$~pF~\cite{Garnier2005}. 
The time constant of the circuit is $\tau_0 \equiv RC = 2.56$~ms. The voltage $V$ across the dipole fluctuates due to Johnson-Nyquist 
thermal noise. We drive the system out of equilibrium by injecting a constant controlled current $I=1.06\cdot10^{-13}$~A. After some $\tau_0$, the 
system is in a NESS.
Electric laws give in this setup~\cite{Garnier2005} :
\begin{equation}
\tau_0 \frac{\dd V}{\dd t} \; + \; V \; = \; RI \; + \; \xi \,,
\label{Electric_equation}
\end{equation}
where $\xi$ is the Gaussian thermal noise, delta-correlated in time of variance $2k_BTR$.

Multiplying (\ref{Electric_equation}) by $V$ and integrating it between $t$ and $t+\tau$, we define the work $W_\tau$, injected 
into the system, and the dissipated heat $Q_{\tau}$, together with the internal energy $U$ :
\begin{eqnarray}
W_\tau & \equiv & \int_t^{t+\tau} V(t')I \dd t'\\
Q_\tau & \equiv & \int_t^{t+\tau} V(t')i_R(t')\dd t' \label{eq:Q_def} \\
\Delta_\tau U & \equiv & \frac{1}{2} C \left( V(t+\tau)^2 - V(t)^2\right) = W_\tau - Q_\tau 
\end{eqnarray}
where $i_R$ is the current flowing in the resistance : $i_R = I - C \frac{\dd U}{\dd t}$. This system has only one degree of freedom,
so the trajectory in phase space is defined by the voltage $V(t)$ alone, and the only external parameter is the constant current $I$. 
So the variation of the "trajectory-dependent" entropy during a time $\tau$ is :
\begin{equation}
\Delta s_{\tau} =  - k_B \ln\left(\frac{p(V(t+\tau))}{p(V(t))}\right)
\label{resist_trajectory_entropy}
\end{equation}

\begin{figure*}[t]
\begin{center}
\includegraphics[width=0.6\linewidth]{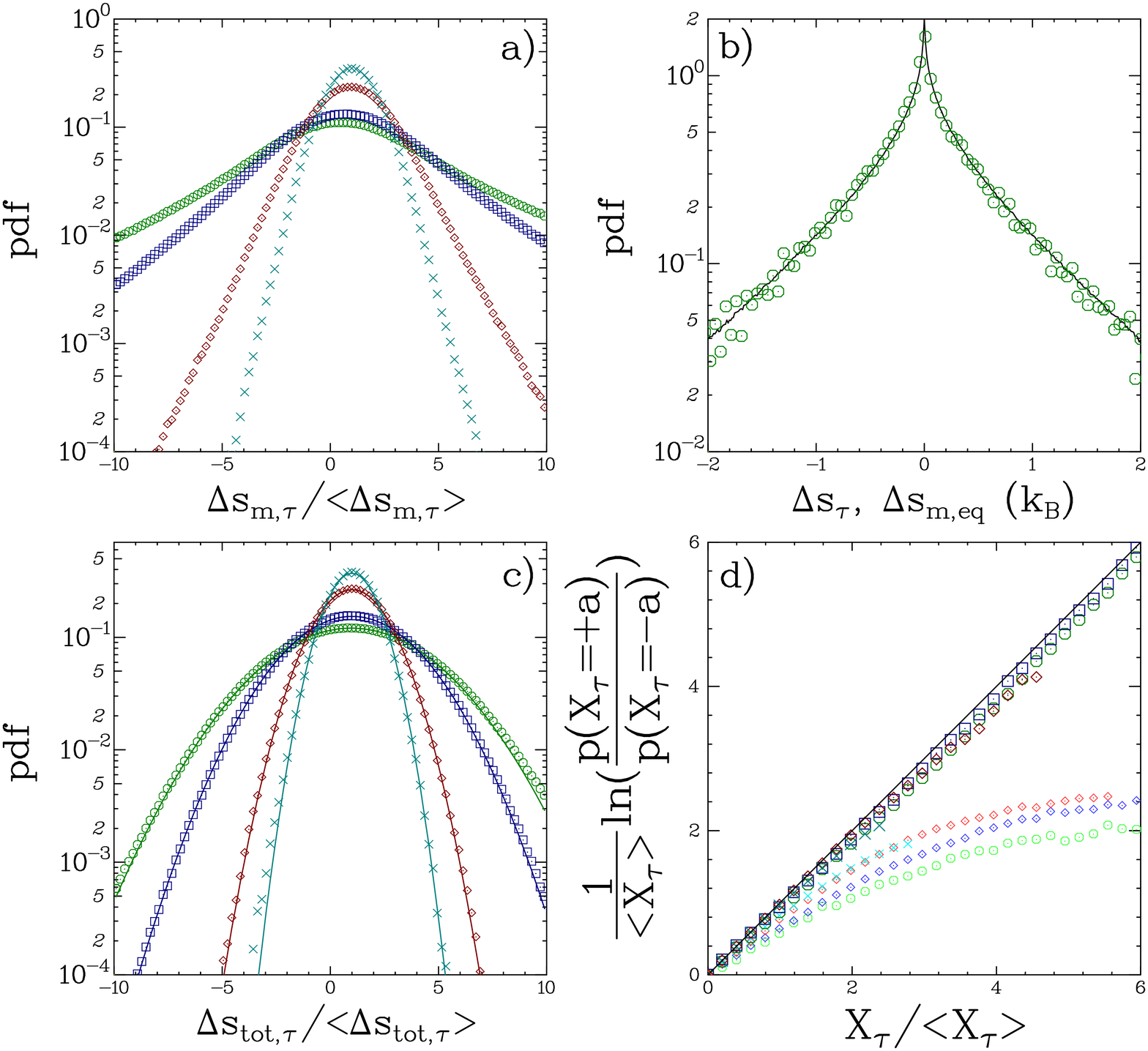}
\caption{Resistance. a) PDFs of the normalized entropy variation in the system $\Delta s_{\rm{m},\tau}/\langle \Delta s_{\rm{m},\tau}\rangle$, with $\tau/\tau_0=2.4$ ($\circ$), $\tau/\tau_0=4.8$ ($\Box$), $\tau/\tau_0=14.5$ ($\diamond$) and $\tau/\tau_0=29$ ($\times$). b) PDF of the variation of "trajectory-dependent" entropy $\Delta s_\tau$ for $\tau/\tau_0=4.8$. The distribution is independent on $\tau/\tau_0$. Continuous line is experimental PDF of the entropy variation at equilibrium ($\Delta s_{\rm{m},\tau,\rm{eq}}$ at $I=0$~A). c)  PDFs of the total entropy production $\Delta s_{\rm{tot},\tau}/\langle \Delta s_{\rm{tot},\tau}\rangle$, with $\tau/\tau_0=2.4$ ($\circ$), $\tau/\tau_0=4.8$ ($\Box$), $\tau/\tau_0=14.5$ ($\diamond$) and $\tau/\tau_0=29$ ($\times$). d) Symmetry functions $\Phi$ for  $\Delta s_{\rm{m},\tau}/\langle \Delta s_{\rm{m},\tau}\rangle$ (small symbols in light colors) and $\Delta s_{\rm{tot},\tau}/\langle \Delta s_{\rm{tot},\tau}\rangle$ (large symbols in dark colors) for the same values of $\tau/\tau_0$.}
\label{fig:resist_PDF}
\end{center}
\end{figure*}

Fluctuation relations for the injected work and the dissipated heat (or entropy change in medium) in this system 
have been reported in~\cite{Garnier2005}. Let us recall the experimental results for the dissipated heat $T.\Delta s_{\rm{m},\tau}$,
for several values of the integration time. Its average value $\langle T.\Delta s_{\rm{m},\tau}\rangle$ is equal to the average of injected work and linear in $\tau$. The PDFs of $T.\Delta s_{\rm{m},\tau}$ are plotted in 
figure~\ref{fig:resist_PDF}a) for four values of $\tau/\tau_0$.  They are 
not Gaussian for small times $\tau$ and extreme events have an exponential 
distribution. We now describe new results in this system. The PDF of the "trajectory-dependent" entropy $\Delta s_{\tau}$ is 
plotted in figure~\ref{fig:resist_PDF}b) ; we have superposed to it the PDF of $\Delta s_{\rm{m},\tau}$ at 
equilibrium ($I=0$~A). These two PDFs are independent of the integration time. The two 
curves match perfectly within experimental errors. Therefore the 
"trajectory-dependent" entropy is equal in this case to the fluctuations of the entropy exchanged with the thermal bath at equilibrium. The average value of this "trajectory-dependent" entropy is zero within experimental errors ; so the total entropy production has the same average value than the entropy $\Delta s_{\rm{m},\tau}$. The PDFs of the total entropy $\Delta s_{\rm{tot},\tau}/\langle \Delta s_{\rm{tot},\tau}\rangle$, computed by adding
$\Delta s_{\tau}$ to $\Delta s_{\rm{m},\tau}$, are plotted in figure~\ref{fig:resist_PDF}c) for different values of $\tau$ ; they are all Gaussian.

In figure~\ref{fig:resist_PDF}d), we have plotted the symmetry functions of the dissipated heat ($T\Delta s_{\rm{m},\tau}$) together with those of the total entropy. $\Phi(\Delta s_{\rm{m},\tau})$ is a non linear function of $\Delta s_{\rm{m},\tau}$. The linearity is recovered for  $\Delta s_{\rm{m},\tau} < \langle \Delta s_{\rm{m},\tau} \rangle$. The slope tends to $1$ for large integration time. Entropy change in medium satisfies relation~(\ref{FT:def})~\cite{Garnier2005}.
On the contrary, the symmetry functions for the total entropy are linear with $\Delta s_{\rm{tot},\tau}$ 
for all integration times $\tau$ and for all values of $\Delta s_{\rm{tot},\tau}$ : $\Phi(\Delta s_{\rm{tot},\tau}) = \Sigma(\tau) \Delta s_{\rm{tot},\tau}$. 
The slope $\Sigma(\tau)$ is equal to $1$ for all integration times within experimental errors. Measurements can be done for other values 
of injected current and we find the same results. Thus total entropy satisfies relation~(\ref{eq:FT_total_entropy}).

This experimental result can be explained using a first order 
Langevin equation and noting that fluctuations of the voltage 
$\delta V(t) = V(t) -\langle V(t) \rangle$, when a current is 
applied, are identical to those at equilibrium~\cite{Garnier2005}. 
Thus the voltage $V(t)$ has a Gaussian distribution with 
mean $\langle V(t)\rangle = R.I$ and variance $(k_B T/C)$,
whereas its autocorrelation function is the same out of equilibrium 
and at equilibrium:
\begin{equation}
\langle \delta V(t+\tau) \delta V(t)\rangle = \frac{k_B T}{C} e^{(-\tau/\tau_0)} \,.
\label{eq:V:autocorr}
\end{equation} 
So we can compute the expression of the "trajectory-dependent" entropy 
from eq.~(\ref{resist_trajectory_entropy}) and we find:
\begin{eqnarray}
T.\Delta s_{\tau} &=& \frac{1}{2}C \delta V(t+\tau)^2 - \frac{1}{2}C \delta V(t)^2
\label{eq:Q_resist}
\end{eqnarray}
Fluctuations of the voltage are identical to those at equilibrium, thus "trajectory-dependent" entropy is equal to the variation of internal energy divided by $T$ at equilibrium and to the opposite of the entropy variation in the system at equilibrium.
Using eq.~(\ref{eq:def_total_entropy}), ~(\ref{eq:Q_def}) and~(\ref{eq:Q_resist}), 
the expression of the total entropy is :
\begin{equation}
T.\Delta s_{\rm{tot},\tau} = I \int_{t}^{t+\tau} V(t') - I \tau_0 \left( \delta V(t+\tau) - \delta V(t)\right)\dd t'
\end{equation}
We experimentally see that the PDF of $\Delta s_{\rm{tot},\tau}$ is Gaussian, so fully characterized by its mean value and its variance. Its average value is linear in 
$\tau$ ($\langle \Delta s_{\rm{tot},\tau} \rangle = R I^2 \tau/T$) and equals the mean injected work divided by the temperature $T$. Its variance is computed using eq.~(\ref{eq:V:autocorr}) and we obtain $\sigma_{\Delta s_{\rm{tot},\tau}}^2 = 2 R I^2 k_B \tau/T$. For a Gaussian distribution, the symmetry function of $\Delta s_{\rm{tot},\tau}$
is linear with $\Delta s_{\rm{tot},\tau}$, that is $\Phi(\Delta s_{\rm{tot},\tau}) = \Sigma \Delta s_{\rm{tot},\tau}$.
The slope $\Sigma$ is related to the mean and the variance of 
the total entropy : $\Sigma = 2\langle \Delta s_{\rm{tot},\tau} \rangle/\sigma_{\Delta s_{\rm{tot},\tau}}^2 = 1/k_B$. 
Experimentally, we have normalized the total entropy by $k_B$, and the slope is equal to $1$ for any $\tau$ as 
shown in fig.~\ref{fig:resist_PDF}d).

We now consider the second experimental system. It is a torsion
pendulum in a viscous fluid which acts as a thermal bath at temperature $T$. 
The system is driven out of equilibrium by an external deterministic 
time dependent torque $M$. All the details of the experimental setup can be found
in~\cite{Joubaud2007}. The torsional stiffness of the oscillator
is $C = 4.65\cdot 10^{-4}$~N.m.rad$^{-1}$,  its viscous damping $\nu$, its total moment of inertia $I_{\mathrm{eff}}$, its resonant frequency 
$f_0 = \omega_0/(2\pi) = \sqrt{C/I}/(2\pi) = 217$~Hz and its relaxation time 
$\tau_\alpha = \nu/(2I_{\mathrm{eff}}) = 9.5$~ms. The angular motion of the oscillator obeys a 
second order Langevin equation :
\begin{equation}
\frac{\dd^2 \theta}{\dd t^2} + \frac{2}{\tau_\alpha} \frac{\dd \theta}{\dd t} + \omega_0^2 \theta = \frac{M+\eta}{C} \,,
\end{equation}
where $\eta$ is the thermal noise, delta-correlated in time of variance 
$2 k_B T \nu$. The work injected into 
the system between $t_i$ and $t_i+\tau$ is :
\begin{equation}
W_\tau =\int_{t_i}^{t_i+\tau} M(t') \frac{\dd \theta}{\dd t}(t') \dd t' \,.
\end{equation}
The dissipated heat is computed according to eq.~(\ref{eq:first_law}) where, in this case, the internal energy $U(t)$ is :
\begin{equation}
U(t) = \frac{1}{2} C \theta(t)^2 + \frac{1}{2} I_{\mathrm{eff}} \dot{\theta}(t)^2
\end{equation}
We investigate a periodic forcing : 
$M(t) = M_0 \sin(\omega_d t)$ ( $M_0 = 0.78$~pN.m and $\omega_d/(2 \pi) = 64$~Hz). 
The integration time $\tau$ is chosen to be a multiple of the period 
of the forcing : $\tau_n = 2 n \pi / \omega_d$. The average responses of the
system $<\theta (t) > $ and $<\dot{\theta} (t) >$  are periodic function of the pulsation  $\omega_d$ and the system is in a steady state. Results for injected work and dissipated heat in this case are reported in 
\cite{Douarche2006, Joubaud2007}.

The "trajectory-dependent" entropy is not as simple as in the case of 
the resistance. The system has two independent degrees of freedom 
($\theta$ and $\dot{\theta}$) and its expression is :
\begin{equation}
\Delta s_{\tau_n}= - k_B \ln \left(\frac{p(\theta(t_i+\tau_n),\varphi).p(\dot{\theta}(t_i+\tau_n,\varphi))}{p(\theta(t_i+\tau_n),\varphi).p(\dot{\theta}(t_i+\tau_n,\varphi)}\right)
\label{pendule_heat_trajectory}
\end{equation}
The DFT (eq.~(\ref{eq:FT_total_entropy})) is valid for each fixed starting phase $\varphi = t_i\omega_d$~\cite{Seifert2007}. For computing correctly the total entropy, we have to calculate the PDFs of the angular position and the angular velocity for each initial phase $\varphi$. Then we compute the "trajectory-dependent" entropy. As fluctuations of $\theta$ and $\dot{\theta}$ are independent of $\varphi$~\cite{Joubaud2007}. These distributions correspond to the equilibrium fluctuations of $\theta$ and $\dot{\theta}$ around the mean trajectory defined by $\langle \theta(t) \rangle$ and $\langle \dot{\theta}(t) \rangle$. As a consequence, we can average $\Delta s_{\tau_n}$ over $\varphi$ which improves a lot the statistical accuracy. We stress that it is not equivalent to calculate first the PDFs over all values of $\varphi$ --- which would correspond here to the convolution of the PDF of the fluctuations with the PDF of a periodic signal --- and then compute the trajectory dependent entropy.

\begin{figure*}[t]
\begin{center}
\includegraphics[width=0.6\linewidth]{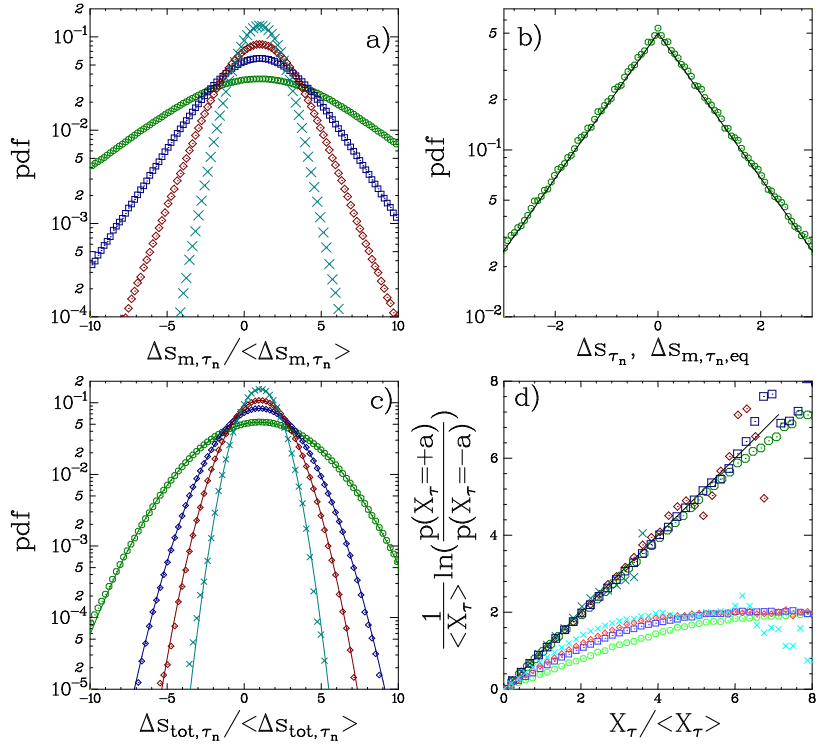}
\caption{Torsion pendulum. 
a) PDFs of the normalized entropy variation $\Delta s_{\rm{m},\tau_n}/\langle \Delta s_{\rm{m},\tau_n}\rangle$ integrated over $n$ periods of forcing, with $n=7$ ($\circ$), $n=15$ ($\Box$), $n=25$ ($\diamond$) and $n=50$ ($\times$). 
b) PDFs of $\Delta s_{\tau_n}$, the distribution is independent of $n$ and here $n=7$. Continuous line is the theoretical prediction for equilibrium entropy exchanged with thermal bath $\Delta s_{\rm{m},\tau_n,eq}$. 
c) PDFs of the normalized total entropy $\Delta s_{\rm{tot},\tau_n}/\langle \Delta s_{\rm{tot},\tau_n}\rangle$, with $n=7$ ($\circ$), $n=15$ ($\Box$), $n=25$ ($\diamond$) and $n=50$ ($\times$). d) Symmetry functions for the normalized entropy variation in the system (small symbols in light colors) and for the normalized total entropy (large symbols in dark colors) for the same values of $n$.}
\label{pendule_sinus:PDF}
\end{center}
\end{figure*}

In figure~\ref{pendule_sinus:PDF}a), we recall the main results 
for the dissipated heat $T\Delta s_{\rm{m},\tau_n}$. Its average value $\langle T.\Delta s_{\rm{m},\tau_n}\rangle$ is linear 
in $\tau_n$ and equal to the injected work. The PDFs of $T.\Delta s_{\rm{m},\tau_n}$ are 
not Gaussian and extreme events have an exponential distribution. Let us now describe new results in this system. The PDF of the "trajectory-dependent" entropy is plotted in 
fig.~\ref{pendule_sinus:PDF}b); it is independent of $n$. We 
superpose to it the PDF of the variation of internal energy divided by $T$
at equilibrium : the two curves match perfectly within 
experimental errors, so the "trajectory-dependent" entropy can again be considered as the entropy exchanged with the thermostat if the system is at equilibrium. 
The average value of $\Delta s_{\tau_n}$ is zero, so the average value of 
the total entropy is equal to the average of injected power divided by $T$. In 
fig.~\ref{pendule_sinus:PDF}c), we plot the PDFs of the normalized 
total entropy for four typical values of integration time. We find 
that the PDFs are Gaussian for any time.

The symmetry functions 
of the dissipated heat $\Phi(\Delta s_{\rm{m},\tau_n})$ 
and the total entropy $\Phi(\Delta s_{\rm{tot},\tau_n})$ are plotted in fig.~\ref{pendule_sinus:PDF}d). 
For the dissipated heat, the symmetry function is a non linear function of  $\Delta S_{m,\tau}$ and we observe a linear behavior for $\Delta s_{\rm{m},\tau_n}<\langle \Delta s_{\rm{m},\tau_n} \rangle$ with a 
slope that tends to $1$ for large time.
For the normalized total entropy, the symmetry functions are linear 
with $\Delta s_{\rm{tot},\tau_n}$ for all values of $\Delta s_{\rm{tot},\tau_n}$ and the slope is 
equal to $1$ for all values of $\tau_n$. Note that it is not exactly the case 
for the first values of $\tau_n$ because these are the times over which the 
statistical errors are the largest and the error in the slope 
is large.

In ref.~\cite{Joubaud2007}, we have already shown that, when a torque is applied, the fluctuations around $\langle \theta(t)\rangle$ have the same statistics and the same 
dynamics as fluctuations at equilibrium. Using the expression of 
the distribution of the angular position and the angular velocity, 
we can compute the "trajectory-dependent" entropy from 
eq.~(\ref{pendule_heat_trajectory}):
\begin{eqnarray}
T.\Delta s_{\tau_n} &=& \frac{1}{2} C \left( \delta \theta(t_i+\tau_n)^2 -\delta \theta(t_i)^2\right) \nonumber\\ &+&\frac{1}{2} I_{\mathrm{eff}} \left(\delta \dot{\theta}(t_i+\tau_n)^2 -\delta \dot{\theta}(t_i)^2\right)
\end{eqnarray}
where $\delta \theta$ and $\delta \dot{\theta}$ are the fluctuations around the mean trajectory : $\theta(t) = \langle \theta(t) \rangle + \delta \theta(t)$ and $\dot{\theta}(t) = \langle \dot{\theta}(t) \rangle + \delta \dot{\theta}(t)$.
The "trajectory-dependent" entropy is equivalent to the variation of internal energy for the system at equilibrium divided by the temperature of the bath. The total entropy is :
\begin{eqnarray}
T.\Delta s_{\rm{tot},\tau_n} &=& W_{\tau_n} - \frac{1}{2} C \bar{\theta}(t_i) (\delta\theta(t_i+\tau_n) - \delta \theta(t_i) \nonumber \\&-& \frac{1}{2} I_{\mathrm{eff}} \dot{\bar{\theta}}(t_i)(\delta \dot{\theta}(t_i+\tau_n) - \delta \dot{\theta}(t_i))
\end{eqnarray}
Experimentally, we observe that the PDF of the total entropy is Gaussian.
We can calculate the mean value and the variance of the total entropy for any $\tau_n$. We find that $\langle T.\Delta s_{\rm{tot},\tau_n} \rangle$ is 
equal to the injected work. 
Using the first 
law of thermodynamics and the equation of motion, we write 
the dissipated heat as the difference between the viscous 
dissipation and the work of the thermal noise :
\begin{equation}
T.\Delta s_{\rm{m},\tau_n} = \int_{t_i}^{t_i+\tau_n} \nu \dot{\theta}(t')^2 \dd t'- \int_{t_i}^{t_i+\tau_n} \dot{\theta}(t') \eta(t') \dd t'
\end{equation}
This expression holds at equilibrium as well as out of equilibrium.
The only difference is that when out of equilibrium 
$\langle \theta(t) \rangle \neq 0$ and $\langle \dot{\theta}(t) \rangle \neq 0$. After some algebra the total entropy is :
\begin{eqnarray}
T.\Delta s_{\rm{tot},\tau_n} &=& \nu \int_{t_i}^{t_i+\tau_n} \langle \dot{\theta}(t') \rangle^2 \dd t' \nonumber\\ &+& 2 \nu \int_{t_i}^{t_i+\tau_n} \langle \dot{\theta}(t') \rangle \delta \dot{\theta}(t')  \dd t'\nonumber\\ &-& \int_{t_i}^{t_i+\tau_n} \langle \dot{\theta}(t') \rangle \eta(t') \dd t'
\end{eqnarray}

The  average value of the total entropy is 
$\nu \int_{t_i}^{t_i+\tau_n} \langle \dot{\theta}(t') \rangle^2 \dd t'$ 
and its variance is :
\begin{eqnarray}
\langle \sigma_{\Delta s_{\rm{tot},\tau_n}}^2 \rangle &=& 2 k_B \langle \Delta s_{\rm{tot},\tau_n}\rangle + B/T^2 \nonumber\\ B&=& 4\nu\int \int \dd u \dd v \langle \dot{\theta}(u) \rangle \langle \dot{\theta}(v) \rangle \psi(u,v)\label{excess_heat_computation}\\
\psi(u,v)&=&\nu \langle \delta \dot{\theta}(u) \delta \dot{\theta}(v) \rangle - \langle \delta \dot{\theta}(u) \eta(v)\rangle
\end{eqnarray}
The first term in the function $\psi$ is the autocorrelation function 
of the angular velocity. This function is symmetric around $u=v$. 
The second term is the correlation function of the angular velocity
 with the noise. Due to the causality principle, this term vanishes 
if $u<v$. Changing variables to $r = (u+v)/2$ and $s = u-v$, 
eq.~(\ref{excess_heat_computation}) can be rewritten :
\begin{eqnarray}
B = \int_{t_i}^{t_i+\tau_n} \dd r \int_{0}^{\tau_n} \langle \dot{\theta}(r+s/2) \rangle \langle \dot{\theta}(r-s/2) 
\rangle\tilde{\psi}(r,s)\\
\tilde{\psi}(r,s) =2 \nu \langle \delta \dot{\theta}(s) \delta \dot{\theta}(0) \rangle - \langle \delta \dot{\theta}(s) \eta(0)\rangle
\end{eqnarray}
After some algebra we can show that the correlation function $\langle\delta \dot{\theta}(s)\eta(0)\rangle$ is two times the autocorrelation 
function $\langle \delta\dot{\theta}(s)\delta\dot{\theta}(0)\rangle$, therefore $\tilde{\psi} = 0$. Thus 
we obtain that the variance of the total entropy is equal to 
$2 k_B \langle \Delta s_{\rm{tot},\tau_n}\rangle$. The Fluctuation Relation for 
the total entropy is : $\Phi(\Delta s_{\rm{tot},n})=1/k_B \Delta s_{\rm{tot},n}$ for all
times $\tau$, for all values of $\Delta s_{\rm{tot},n}$ and for any kind 
of stationary forcing.

For our two experimental systems, we have obtained that the  "trajectory-dependent" entropy can be considered as the entropy variation in the system in a time $\tau$ that
one would have if the system was at equilibrium. Therefore the 
total entropy is the additional entropy due to the presence of the 
external forcing : this is the part of entropy which is created due to the non-equilibrium stationary process. The derivation we gave 
for a second order Langevin equation can be extended to a first
order Langevin equation : total entropy (or excess entropy) satisfies
the Fluctuation Theorem for all times and for all kind of stationary 
intensity injected in the circuit or all kind of stationary external 
torque. The ratio between 
the average value of the total entropy at $\tau = \tau_\alpha$ (relaxation time) and the variance $\sigma_{\rm{eq}}^2$ of fluctuations the entropy variation at equilibrium characterizes the distance from equilibrium $d$ :
\begin{equation}
d^2 = \frac{\langle \Delta s_{\rm{tot},\tau_\alpha} \rangle}{\sigma_{\rm{eq}}} = \sqrt{\frac{3 n_d}{8}}\frac{\sigma_{\Delta s_{\rm{tot},\tau_\alpha}}^2}{\sigma_{\rm{eq}}^2} 
\label{dist_from_eq}
\end{equation}
where $n_d$ is the number of degrees of freedom. For the second equality, we use the Gaussianity of $\Delta s_{\rm{tot},\tau}$, {\em i.e.} $\sigma_{\Delta s_{\rm{tot},\tau}}^2 = 2 k_B  \langle \Delta s_{\rm{tot},\tau} \rangle$ and the variance of of the fluctuation of entropy variation at equilibrium in terms of $k_B$, {\em{i.e.}} $\sigma_{\rm{eq}}^2 = 3/2 n_d k_B^2$. It turns out that this expression for $d$ is the same that was defined in~\cite{Joubaud2007} with a completely different approach. Eq.~(\ref{dist_from_eq}) indicates that, when the system is driven far from equilibrium, $\sigma_{\rm{eq}}$ becomes negligible. As a consequence the fluctuations of the total entropy become equal to the fluctuations of the entropy variation in the system in a time $\tau$. In other words, the PDFs of the dissipated heat far from equilibrium are Gaussian.

In conclusion, we have studied the total entropy when the system is driven in a non-equilibrium steady state. We have shown that the Fluctuation Theorem for the total entropy is valid not only in the limit of large times, as it is the case for injected work and dissipated heat, but also for all integration times and all fluctuation amplitudes. In the two examples we have discussed, the total entropy corresponds to the difference between the entropy change in medium out of equilibrium and at equilibrium.

We thank U. Seifert and K. Gawedzki for useful discussions. 
This work has been partially supported by ANR-05-BLAN-0105-01.

\end{document}